\documentclass[12pt]{article}
\usepackage[T1]{fontenc}
\usepackage{graphics}

\makeatletter

\providecommand{\LyX}{L\kern-.1667em\lower.25em\hbox{Y}\kern-.125emX\@}

\usepackage{a4}
\usepackage{cite}
\input{epsfig.sty}
\def\fnum@table{\tablename~{\bf\thetable}}
\def\fnum@figure{\figurename~{\bf\thefigure}}
\def\tablename{\footnotesize{\bf Table}}
\def\figurename{\footnotesize{\bf Figure}}

\makeatother

\begin{document}

\title{\textbf{\Huge Consistent Treatment of Soft and Hard }\\
\textbf{\Huge Processes in Hadronic Interactions}\\
\Huge }

\author{{\small F.M. Liu\protect\( ^{1,2}\protect \), H.J. Drescher\protect\( ^{3}\protect \),
S. Ostapchenko\protect\( ^{4,5}\protect \), T. Pierog\protect\( ^{1}\protect \),
and K. Werner\protect\( ^{1}\protect \)}\textbf{\emph{\small }}\\
\\
 \textit{\small \protect\( ^{1}\protect \) SUBATECH, Université de Nantes --
IN2P3/CNRS -- Ecole des Mines,  Nantes, France }\\
\textit{\small \protect\( ^{2}\protect \)Institute of Particle Physics , Huazhong
Normal University, Wuhan, CHINA}\\
 \textit{\small \protect\( ^{3}\protect \) Physics Department, New York University,
New York, USA} {\small }\textit{\small }\\
\textit{\small \protect\( ^{4}\protect \) Institut für Kernphysik, Forschungszentrum
Karlsruhe, D-76021 Karlsruhe, Germany} {\small }\textit{\small }\\
\textit{\small \protect\( ^{5}\protect \) Moscow State University, Institute
of Nuclear Physics, Moscow, Russia }\\
 \textit{\small }\small }

\maketitle
\begin{abstract}
The QCD improved parton model is a very successful concept to treat processes
in hadronic interactions, whenever large partonic transverse momenta \( p_{\bot } \)
are involved. However, cross sections diverge in the limit \( p_{\bot }\rightarrow 0 \),
and the usual treatment is the definition of a lower cutoff \( p_{\bot \mathrm{min}} \),
such that processes with a smaller \( p_{\bot } \) -- so-called soft processes
-- are simply ignored, which is certainly not correct for example at RHIC energies.
A more consistent procedure amounts to introduce a technical parameter \( Q^{2}_{0} \),
referred to as soft virtuality scale, which is nothing but an artificial borderline
between soft and hard physics. We will discuss such a formalism, which coincides
with the improved parton model for high \( p_{\bot } \) processes and with
the phenomenological treatment of soft scattering, when only small virtualities
are involved. The most important aspect of our approach is that it allows to
obtain a smooth transition between soft and hard scattering, and therefore no
artificial dependence on a cutoff parameter should appear.
\end{abstract}

\section{Introduction}

The standard parton model approach to hadron-hadron scattering amounts to presenting
the partons of projectile and target by momentum distribution functions, \( f_{i} \)
and \( f_{j} \), and calculating inclusive cross sections for the production
of parton jets with the squared transverse momentum \( p_{\perp }^{2} \) larger
than some cutoff \( p^{2}_{\bot \mathrm{min}} \) as
\begin{equation}
\sigma _{\mathrm{incl}}\! (s)=\sum _{ij}\int _{p^{2}_{\bot }>p^{2}_{\bot \mathrm{min}}}dp_{\perp }^{2}dx^{+}dx^{-}f_{i}(x^{+},p_{\perp }^{2})f_{j}(x^{-},p_{\perp }^{2})\, \, d\hat{\sigma }_{ij}/dp^{2}_{\bot }(x^{+}x^{-}s)
\end{equation}
 where \( d\hat{\sigma }_{ij}/dp_{\perp }^{2} \) is the elementary parton-parton
cross section and \( i,j \) represent parton flavors. Many Monte Carlo applications
are based on the above formula: ISAJET \cite{pai86}, PYTHIA \cite{ben87,sjo87},HERWIG
\cite{web84,mar84,mar88}, HIJING \cite{wan96} -- for more details see the
review \cite{sjo89}. 

It is the cutoff \( p^{2}_{\bot \mathrm{min}} \) which prevents the above integral
to diverge, and there are several justifications for using such a cutoff. If
one is only interested in hard processes, for example to study large \( p_{\bot } \)
jets, such a cutoff can be used without problems, since soft reactions are simply
not considered. 

The situation is much more complicated when it comes to treating an average
event, since here the soft component plays an important role. In nuclear scattering,
one may argue with screening effects: in heavy nuclei at very high energies,
the parton density will be reduced due to screening, and this may be effectively
done via a cutoff. Real life is more complicated: we have finite nuclei at finite
energy, so soft physics still plays a role and has to be treated properly. The
same is true for screening corrections: in particular for Monte Carlo applications
one cannot treat them just introducing a cutoff or modifying structure functions,
this is only valid for calculating inclusive spectra based on the assumption
that factorization holds. 

There is already some indication that the above-mentioned concept comes to its
limits: taking parton distribution functions compatible with latest HERA measurements,
it seems to be impossible to get the correct energy dependence of the proton-proton
cross section, which can only be cured by an unreasonable assumption like an
energy dependent cutoff(see, for example,\cite{boo94}).

We think that one should well separate the aspect soft/hard physics and screening.
One may introduce a soft scale \( Q^{2}_{0} \), which is meant to be the borderline
between soft and hard physics, such that above this scale one may use perturbative
methods, whereas below one has to work with parameterizations. We do not think
that the physics changes abruptly when crossing this artificial border, it is
more a technical problem that we know how to do calculations only above the
scale \( Q^{2}_{0} \). It is not at all a serious problem that one has to rely
on a phenomenological treatment at low virtualities, since there exists a wealth
of information about this topic, in particular since such concepts as the Pomeron
can be investigated at the HERA collider. Being just a borderline between soft
and hard physics, there is absolutely no reason that the parameter \( Q^{2}_{0} \)
should be energy dependent, or vise versa - introducing an energy dependent
cutoff necessarily implies changing physics content of the ``soft'' part of
the interaction as with the increasing energy harder and harder partons are
treated as being ``soft''. 

Naturally, at high energies the role of screening corrections due to so-called
enhanced diagrams becomes extremely important. Our approach allows to treat
corresponding contributions explicitely, as coming from Pomeron-Pomeron interactions
in the ``soft'' nonperturbative region. This allows to achieve a microscopic
description of the interaction process, to resolve the seeming contradiction
between the energy dependence of hadronic interaction cross sections and the
realistic structure functions measured by HERA, and to get rid of the artificial
dependence of the results on the technical parameter \( Q^{2}_{0} \) \cite{ost01}.

\section{Qualitative Discussion of Nucleon-Nucleon Scattering}

\subsection{The Structure of the Nucleon}

\begin{figure}[htb]
{\par\centering \resizebox*{!}{0.15\textheight}{\includegraphics{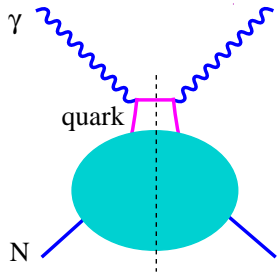}} \par}

\caption{A virtual photon probing the structure of the nucleon. \label{fig:struc1}}
\end{figure}

Nucleons are composite objects, and high energy nucleon-nucleon scattering requires
therefore some discussion on the intrinsic structure of the nucleon. Let us
consider deep inelastic scattering, where a virtual photon probes the nucleon
structure, see fig. \ref{fig:struc1}, where we show a cut diagram, corresponding
to a squared amplitude.
 As shown in the figure, the photon couples to a quark from the nucleon, allowing
this way the measurement of the quark momentum distributions (nucleon structure
functions). For photons with large virtualities, one may apply the methods of
perturbative QCD in order to understand the structure of the ``blob'' in the
above figure: we know that the final quark is emitted from a parton, which itself
is emitted from another parton and so on. Employing the leading logarithmic
approximation one obtains the following picture (see fig. \ref{fig:struc2}):
\begin{figure}[htb]
{\par\centering \resizebox*{!}{0.15\textheight}{\includegraphics{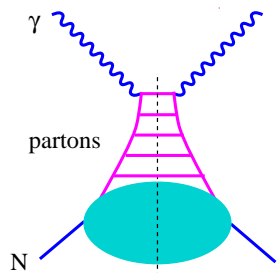}} \par}

\caption{A highly virtual photon couples to the ``end'' of a parton ladder, emitted
from the nucleon. \label{fig:struc2}}
\end{figure}
 The nucleon emits a parton with a virtuality of the order of some soft scale
\( Q_{0}^{2} \) and with some light cone momentum fraction \( x \), which
emits itself parton with higher virtuality but smaller \( x \) and so on, such
that the last parton couples to the photon. It is a question of taste to consider
the parton ladder of being a part of the internal structure of the nucleon and
the photon to interact with a quark of the nucleon, or to consider only the
first parton at scale \( Q_{0}^{2} \) (the ``soft'' parton) being a constituent
of the nucleon, whereas the parton ladder is considered to be a part of the
interaction. Concerning the soft partons, one distinguishes between sea quarks
or gluons and valence quarks. Whereas the latter ones have typical light cone
momentum fraction distributions of the form \( x^{-0.5} \), the former ones
are distributed as \( x^{-\alpha _{0}} \), with \( \alpha _{0} \) being at
least unity. This means that sea quarks or gluons have typically very small
\( x \)-values, contrary to the valence quarks. What is the mass of the ``blob''
in fig. \ref{fig:struc2}, left behind the emitted quark? Having emitted a parton
with momentum \( k \) and virtuality \( k^{2}=-Q_{0}^{2} \) from the nucleon,
considered forward moving with momentum \( p \), we obtain in the high energy
limit for the squared mass of the remainder (nucleon minus parton)
\[
\hat{s}=-p^{+}k^{-}=p^{+}Q_{0}^{2}/k^{+}=Q_{0}^{2}/x,\]
where \( p^{\pm }=p_{0}\pm p_{3},\, k^{\pm }=k_{0}\pm k_{3} \) are the light
cone momenta of the nucleon and of the emitted parton correspondingly and \( x=k^{+}/p^{+} \)
is the light cone momentum fraction of the parton. This shows that in case of
small \( x \) (sea quarks or gluons), the remainder has a large mass; whereas
for valence quarks, we obtain masses of the order of \( Q_{0} \). This means
sea quarks or gluons are not elementary nucleon constituents, but they are rather
emitted from some object with large mass and low virtuality \cite{don94}. Such
objects are usually referred to as soft Pomerons, and very powerful phenomenological
techniques exist to deal with them.

\begin{figure}[htb]
{\par\centering \resizebox*{!}{0.15\textheight}{\includegraphics{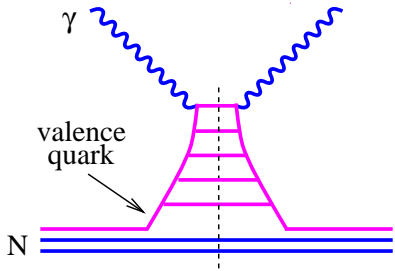}} \par}

\caption{Photon-nucleon scattering, with the first (soft) parton being a valence quark.
\label{fig:struc3}}
\end{figure}

\begin{figure}[htb]
{\par\centering \resizebox*{!}{0.15\textheight}{\includegraphics{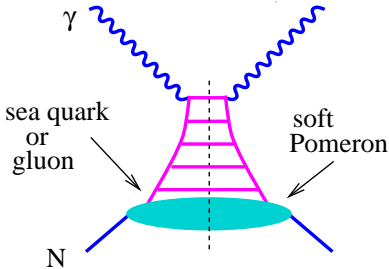}} \par}

\caption{Photon-nucleon scattering, with the first (soft) parton being a sea-quark or
gluon. \label{fig:struc4}}
\end{figure}

So we distinguish between two contributions, depending on whether the first
(soft) parton is a sea quark or gluon or a valence quark. In case of valence
quarks, we consider the quark to be a direct constituent of the nucleon, and
the corresponding photon-nucleon scattering diagram is shown in fig. \ref{fig:struc3}.
So one may assume some initial distribution \( F_{\mathrm{val}}(x) \) for the
valence quarks, and treat then the scattering of this valence quark with the
virtual photon. In case of sea quarks or gluons, there has to be a soft Pomeron
between the nucleon and the first parton, where we do not know the precise microscopic
structure, but we know how to parameterize the corresponding amplitudes. So
our graphical representation will be simply a ``blob'', as shown in fig. \ref{fig:struc4}.
This blob stands for a soft Pomeron, and we will use later the corresponding
parameterizations. We do not know how the soft Pomeron couples to the nucleon,
but we may imagine ``some constituent'' carrying some fraction \( x \) of
the nucleons momentum according to some distribution \( F(x) \), which emits
a soft Pomeron, which by itself emits the first perturbative \textbf{\textit{(\( k^{2}\sim -Q_{0}^{2} \))}}
parton. In this sense, the external lower line in fig. \ref{fig:struc4} may
be considered to be such a constituent. 

To summarize: we consider in any case the scattering of a virtual photon with
a nucleon constituent carrying some fraction \( x \) of the nucleons momentum.
In case of valence quarks, this constituent is simply a valence quark with distribution
\( F_{\mathrm{val}}(x) \), in case of sea quarks or gluons, the constituent
is not known, and we use some function \( F(x) \), to be specified later. In
case of valence quarks, the interaction process is treated entirely in the framework
of perturbative QCD, using the leading logarithmic approximation; in case of
sea quarks or gluons, one has in addition to consider a soft Pomeron.

\subsection{Nucleon-Nucleon Scattering }

\begin{figure}[htb]
{\par\centering \resizebox*{!}{0.18\textheight}{\includegraphics{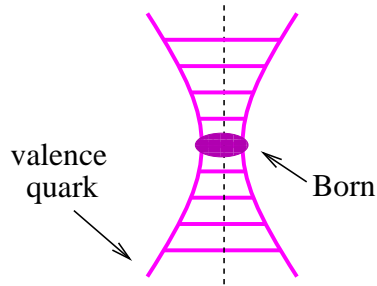}} \par}

\caption{The valence-valence contribution. \label{fig:val-val}}
\end{figure}

Virtual photon-nucleon scattering may be seen as an interaction between the
virtual photon and a nucleon constituent, where the interaction is realized
via the exchange of some composite object -- as discussed in the preceding section.
This exchange object is either a parton ladder (in case of valence quarks involved)
or a parton ladder plus a soft Pomeron (in case of sea quarks or gluons involved).
We easily generalize to nucleon-nucleon scattering. Let us assume that there
are hard partons involved -- with virtualities bigger than the soft scale \( Q_{0}^{2} \).
Employing the leading logarithmic approximation of perturbative QCD, we expect
some hard scattering of two partons with high virtualities, these two parton
being emitted from partons with smaller virtualities and so on, till one reaches
on both ends the soft scale \( Q_{0}^{2} \). So now we have two initial (low
virtuality) partons on both sides, and for either of them we can literally repeat
the discussion of the last chapter: each parton may be of sea or valence type,
and correspondingly we have four contributions. If both partons are of valence
type, we have an interaction between two valence quarks via the exchange of
a parton ladder, as shown in fig. \ref{fig:val-val} (valence-valence contribution).
We do not show in the figure the spectator quarks. The fact that the ladder
rungs getting narrower towards the middle of ladder indicates symbolically virtuality
ordering: a bigger ladder rung represents smaller virtuality. Each of the two
ordered ladders corresponds exactly to the parton ladder employed in deep inelastic
scattering. 
The only new aspect here is the fact that the two ordered ladders are ``glued''
together by means of an elementary Born scattering diagram, which represents
the process with the highest virtuality involved. Since only hard partons contribute,
we refer to the valence-valence contributions also as the ``hard'' one.

\begin{figure}[htb]
{\par\centering \resizebox*{!}{0.18\textheight}{\includegraphics{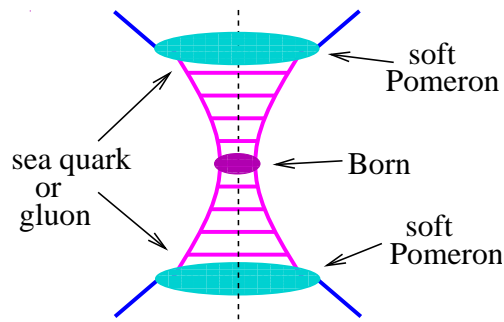}} \par}

\caption{The sea-sea contribution. \label{fig:sea-sea}}
\end{figure}

\begin{figure}[htb]
{\par\centering \resizebox*{!}{0.18\textheight}{\includegraphics{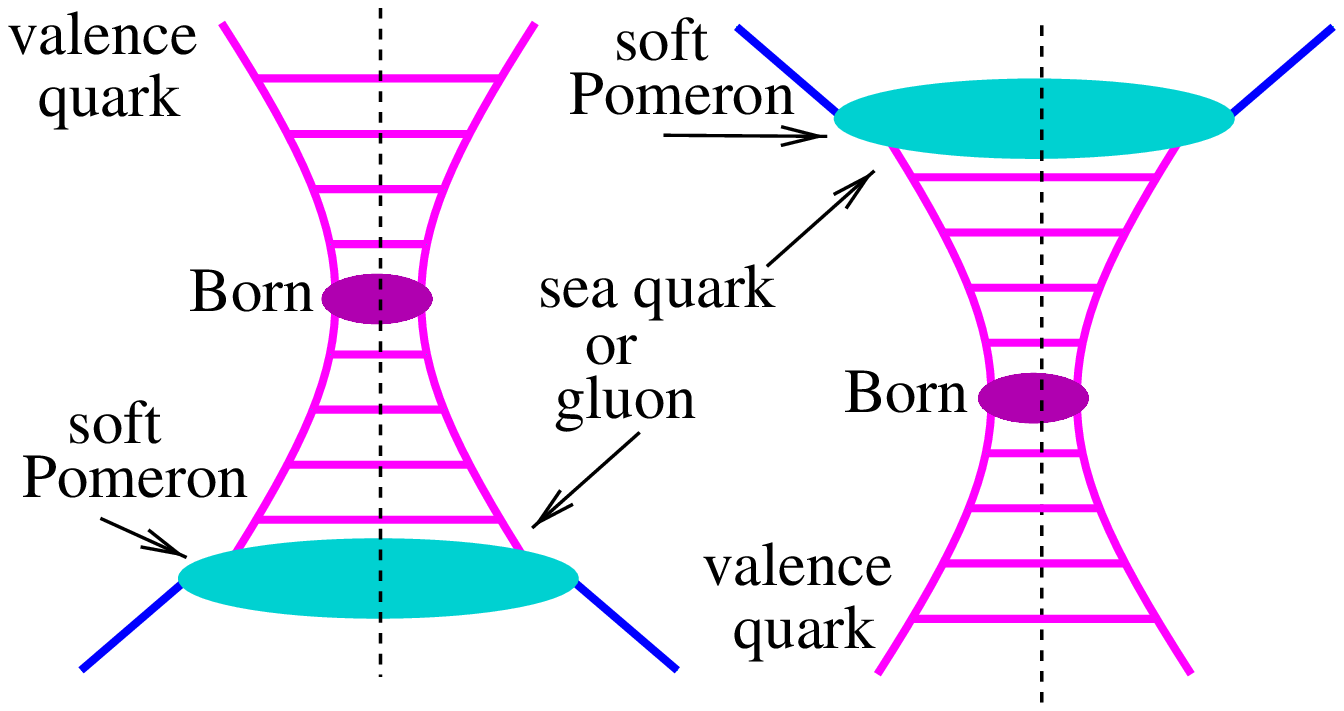}} \par}

\caption{Two ``mixed'' contributions.\label{fig:mixed}}
\end{figure}

We obtain a second contribution, when we consider the scattering of sea quarks
or gluons, as shown in fig. \ref{fig:sea-sea}.
As discussed in the preceding section, there is in general a soft, large mass
object between the sea quark (gluon) \textbf{\textit{}}and the nucleon, which
we identify with the soft Pomeron. So the sea quarks or gluons are internal
lines in the above diagram. The external lines in the figure refer to ``nucleon
constituents'', the precise nature of which we do not need to specify at this
point. We simply assume that they carry a fraction \( x \) of the nucleons
momentum, according to some distribution \( F(x) \). We have of course also
the mixed cases, where a valence quark scatters off a sea quark or gluon, as
shown in fig. \ref{fig:mixed}.
The external legs here are a valence quark on one side and a nucleon constituent
on the other side, since the sea quark or gluon is as usual an internal line.

\begin{figure}[htb]
{\par\centering \resizebox*{!}{0.1\textheight}{\includegraphics{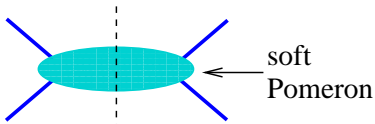}} \par}

\caption{The soft contribution. \label{fig:soft}}
\end{figure}

Let us come back to the sea-sea contribution. Here we consider hard perturbative
partons in the ladder and ``soft partons'' in the shaded area before it, where
the latter ones are not treated explicitly but rather parameterized as soft
Pomeron contributions. One may imagine the case where the hardest parton in
the process \textbf{\textit{}}is already a soft one, with the corresponding
virtuality being smaller than \textbf{\textit{\( Q_{0}^{2} \)}} cutoff\textbf{\textit{,}}
so the hard piece in the middle is reduced to zero. Since this is not possible
for our sea-sea contribution, where one requires always at least one hard parton
in the diagram, one has to add explicitly a purely soft contribution, which
is simply parameterized as a soft Pomeron and graphically represented as a ``blob'',
see fig. \ref{fig:soft}.
We use of course the same parameterization as for the soft Pomerons appearing
in the sea-sea and the mixed contributions, as we are going to discuss later.
The external legs for the soft contribution are nucleon constituents, as for
the sea-sea case.

A complete elementary interaction in hadron-hadron scattering is therefore the
sum of all the above-mentioned contributions: a soft one, a hard one, and three
``semi-hard'' contributions (sea-sea and mixed). We have a smooth transition
from hard to soft physics: lowering the energy will reduce the relative weight
of the semi-hard contributions till they finally die out and we are left with
a purely soft contribution. In our approach, we do not consider soft and hard
physics as completely different: the soft domain is just the continuation of
perturbative domain into a region, where we do not have the technical abilities
to perform rigorous calculations. But physics changes smoothly.

In the following section, we are going to treat the different contributions
in detail.

\section{Detailed Treatment of Nucleon-Nucleon Scattering}

\subsection{Soft Scattering}

Let us first consider the purely soft contribution of the fig.\ \ref{fig:soft},
where all virtual partons appearing in the internal structure of the diagram
have restricted virtualities \( Q^{2}<Q^{2}_{0} \), where \( Q_{0}^{2}\sim 1 \)
GeV\( ^{2} \) is a reasonable cutoff for perturbative QCD being applicable.
Such soft non-perturbative dynamics is known to dominate hadron-hadron interactions
at not too high energies. Lacking methods to calculate this contribution from
first principles, it is simply parameterized and graphically represented as
a `blob', see fig.\ \ref{fig:soft}. It is traditionally assumed to correspond
to multi-peripheral production of partons (and final hadrons) \cite{afs62}
and is described by the phenomenological soft Pomeron exchange contribution.
The scattering amplitude is given as \cite{gri68}
\begin{equation}
\label{tsoft}
T_{\mathrm{soft}}\! \! \left( \hat{s},t\right) =8\pi s_{0}\eta (t)\, \gamma _{\mathrm{part}}^{2}\, \left( \frac{\hat{s}}{s_{0}}\right) ^{\alpha _{\mathrm{soft}}\! (0)}\exp \! \left( \lambda ^{(2)}_{\mathrm{soft}}\! \left( \hat{s}/s_{0}\right) t\right) ,
\end{equation}
with
\begin{equation}
\label{x}
\lambda ^{(n)}_{\mathrm{soft}}\! (z)=nR_{\mathrm{part}}^{2}+\alpha '\! _{\mathrm{soft}}\ln \! z,
\end{equation}
where \( \hat{s}=(p+p')^{2} \)\textbf{\textit{, \( p,p' \)}} are the 4-momenta
of the initial constituent partons in the process before and after the scattering
(considered here as being nearly real, i.e. \textbf{\textit{\( p^{2}=p'^{2}\simeq 0 \))}}.
The parameters \( \alpha _{\mathrm{soft}}\! (0) \), \( \alpha '\! _{\mathrm{soft}} \)
are the intercept and the slope of the Pomeron trajectory, \( \gamma _{\mathrm{part}} \)
and \( R_{\mathrm{part}}^{2} \) are the vertex value and the slope for the
Pomeron-parton coupling, and \( s_{0}\simeq 1 \) GeV\( ^{2} \) is the characteristic
hadronic mass scale. The so-called signature factor \( \eta  \) is given as
\begin{equation}
\label{x}
\eta (t)=i-\cot \frac{\pi \alpha _{_{\rm {P}}}\! (t)}{2}\simeq i.
\end{equation}
 Cutting the diagram corresponds to the summation over multi-peripheral intermediate
hadronic states, connected via unitarity to the imaginary part of the amplitude
(\ref{tsoft}),
\begin{eqnarray}
\frac{1}{i}\mathrm{disc}_{\hat{s}}\, T_{\mathrm{soft}}\! \! \left( \hat{s},t\right)  & = & \! \! \frac{1}{i}\left[ T_{\mathrm{soft}}\! \! \left( \hat{s}\! +\! i0,t\right) -T_{\mathrm{soft}}\! \! \left( \hat{s}\! -\! i0,t\right) \right] \\
 & = & 2\mathrm{Im}\, T_{\mathrm{soft}}\! \! \left( \hat{s},t\right) \\
 & = & \sum _{n,\mathrm{spins},...}\int d\tau _{n}T_{p,p'\rightarrow X_{n}}T^{*}_{\tilde{p},\tilde{p}'\rightarrow X_{n}},\label{x} 
\end{eqnarray}
where \( T_{p,p'\rightarrow X_{n}} \) is the amplitude for the transition of
the initial partons \( p,p' \) into the \( n \)-particle state \( X_{n} \),
\( d\tau _{n} \) is the invariant phase space volume for the \( n \)-particle
state \( X_{n} \) and the summation is done over the number of particles \( n \)
and over their spins and species, the averaging over initial parton colors and
spins is assumed; \( \mathrm{disc}_{\hat{s}}\, T_{\mathrm{soft}}\! \! \left( \hat{s},t\right)  \)
denotes the discontinuity of the amplitude \( T_{\mathrm{soft}}\! \left( \hat{s},t\right)  \)
on the right-hand cut in the variable \( \hat{s} \).

The corresponding profile function for parton-parton interaction, defined as
twice the imaginary part of the Fourier transform \( \tilde{T} \) of \( T \),
divided by the initial parton flux \( 2\hat{s} \),
\begin{equation}
\label{x}
D_{\mathrm{soft}}\! \left( \hat{s},b\right) =\frac{1}{2\hat{s}}2\mathrm{Im}\tilde{T}_{\mathrm{soft}}(\hat{s},b),
\end{equation}
is given as
\begin{eqnarray}
D_{\mathrm{soft}}(\hat{s},b) & = & \frac{1}{8\pi ^{2}\hat{s}}\int \! \! d^{2}q_{\perp }\exp \! \left( -i\vec{q}_{\perp }\vec{b}\right) \, 2\mathrm{Im}\, T_{\mathrm{soft}}\! \left( \hat{s},-q^{2}_{\perp }\right) \\
 & = & \! \frac{2\gamma _{\mathrm{part}}^{2}}{\lambda ^{\! (2)}_{\mathrm{soft}}\! (\hat{s}/s_{0})}\! \! \left( \frac{\hat{s}}{s_{0}}\right) ^{\alpha _{\mathrm{soft}}\! (0)-1}\! \! \! \exp \! \left( \! \! -\frac{b^{2}}{4\lambda ^{\! (2)}_{\mathrm{soft}}\! (\hat{s}/s_{0})}\right) \label{soft d} 
\end{eqnarray}
 For \( t=0 \) one obtains via the optical theorem the contribution \( \sigma _{\mathrm{soft}} \)
of the soft Pomeron exchange to the total parton interaction cross section,
\begin{eqnarray}
\sigma _{\mathrm{soft}}\! \left( \hat{s}\right)  & = & \frac{1}{2\hat{s}}2\mathrm{Im}\, T_{\mathrm{soft}}\! \! \left( \hat{s},0\right) \\
 & = & \int d^{2}bD_{\mathrm{soft}}(\hat{s},b)\\
 & = & 8\pi \gamma _{\mathrm{part}}^{2}\left( \frac{\hat{s}}{s_{0}}\right) ^{\alpha _{\mathrm{soft}}\! (0)-1}.
\end{eqnarray}
 The external legs of the diagram of fig.\ \ref{fig:soft} are ``partonic constituents'',
as discussed in the preceding section.

\subsection{Hard Scattering}

Let us now consider the hard (or valence-valence) contribution, see fig.\ref{fig:val-val}.
All the processes in the interaction diagram are perturbative, i.e. all internal
intermediate partons are characterized by large virtualities \( Q^{2}>Q^{2}_{0} \).
In that case, the corresponding hard parton-parton scattering amplitude \( T^{jk}_{\mathrm{hard}}\! \! \left( \hat{s},t\right)  \)
(\( j,k \) denote the types (flavors) of the initial quarks) can be calculated
using the perturbative QCD techniques \cite{alt82,rey81}, and the intermediate
states contributing to the absorptive part of the amplitude can be defined in
the parton basis. In the leading logarithmic approximation of QCD, summing up
terms where each (small) running QCD coupling constant \( \alpha _{s}(Q^{2}) \)
appears together with a large logarithm \( \ln (Q^{2}/\lambda ^{2}_{\mathrm{QCD}}) \)
(with \( \lambda _{QCD} \) being the infrared QCD scale), and making use of
the factorization hypothesis, one obtains the contribution of the corresponding
cut diagram for \( t=q^{2}=0 \) as the cut parton ladder cross section \( \sigma _{\mathrm{hard}}^{jk}(\hat{s},Q_{0}^{2}) \) \footnote{
Strictly speaking, one obtains the ladder representation for the process only
using axial gauge.
}, which will correspond to the cut diagram, where all horizontal rungs are the
final (on-shell) partons and the virtualities of the virtual \( t \)-channel
partons increase from the ends of the ladder towards the largest momentum transfer
parton-parton process (indicated symbolically by the `blob' in the middle of
the ladder):

\begin{eqnarray}
\sigma _{\mathrm{hard}}^{jk}(\hat{s},Q_{0}^{2}) & = & \frac{1}{2\hat{s}}2\mathrm{Im}\, T^{jk}_{\mathrm{hard}}\! \! \left( \hat{s},t=0\right) \\
 & = & K\, \sum _{ml}\int dx_{B}^{+}dx_{B}^{-}dp_{\bot }^{2}{d\sigma _{\mathrm{Born}}^{ml}\over dp_{\bot }^{2}}(x_{B}^{+}x_{B}^{-}\hat{s},p_{\bot }^{2})\nonumber \\
 & \times  & E_{\mathrm{QCD}}^{jm}(x_{B}^{+},Q_{0}^{2},M_{F}^{2})\, E_{\mathrm{QCD}}^{kl}(x_{B}^{-},Q_{0}^{2},M_{F}^{2})\theta \! \left( M_{F}^{2}-Q^{2}_{0}\right) .\label{sig-jk-hard} 
\end{eqnarray}
 Here \( d\sigma _{\mathrm{Born}}^{ml}/dp_{\bot }^{2} \) is the differential
\( 2\rightarrow 2 \) parton scattering cross section, \( p_{\bot }^{2} \)
is the parton transverse momentum in the hard process, \( m,l \) and \( x_{B}^{\pm } \)
are correspondingly the types and the shares of the light cone momenta of the
partons participating in the hard process, and \( M_{F}^{2} \) is the factorization
scale for the process (we use \( M_{F}^{2}=p^{2}_{\perp }/4 \)). The `evolution
function' \( E^{jm}_{\mathrm{QCD}}(Q^{2}_{0},M_{F}^{2},z) \) represents the
evolution of a parton cascade from the scale \( Q_{0}^{2} \) to \( M_{F}^{2} \),
i.e.\ it gives the number density of partons of type \( m \) with the momentum
share \( z \) at the virtuality scale \( M_{F}^{2} \), resulted from the evolution
of the initial parton \( j \), taken at the virtuality scale \( Q_{0}^{2} \).
The evolution function satisfies the usual DGLAP equation \cite{alt77} with
the initial condition \( E^{jm}_{\mathrm{QCD}}(Q^{2}_{0},Q_{0}^{2},z)=\delta ^{j}_{m}\; \delta (1-z) \).
The factor \( K\simeq 1.5 \) takes effectively into account higher order QCD
corrections.

In the following we shall need to know the contribution of the uncut parton
ladder \( T_{\mathrm{hard}}^{jk}(\hat{s},t) \) with some momentum transfer
\( q \) along the ladder (with \( t=q^{2} \)). The behavior of the corresponding
amplitudes was studied in \cite{lip86} in the leading logarithmic(\( 1/x \))
approximation of QCD. The precise form of the corresponding amplitude is not
important for our application; we just use some of the results of \cite{lip86},
namely that one can neglect the real part of this amplitude and that it is nearly
independent on \( t \), i.e. that the slope of the hard interaction \( R_{\mathrm{hard}}^{2} \)
is negligibly small, i.e. compared to the soft Pomeron slope one has \( R_{\mathrm{hard}}^{2}\simeq 0 \).
So we parameterize \( T_{\mathrm{hard}}^{jk}(\hat{s},t) \) in the region of
small \( t \) as \cite{rys92}
\begin{equation}
\label{t-ladder}
T_{\mathrm{hard}}^{jk}(\hat{s},t)=i\hat{s}\, \sigma _{\mathrm{hard}}^{jk}(\hat{s},Q_{0}^{2})\: \exp \left( R_{\mathrm{hard}}^{2}\, t\right) 
\end{equation}

The corresponding profile function is obtained by calculating the Fourier transform
\( \tilde{T}_{\mathrm{hard}} \) of \( T_{\mathrm{hard}} \) and dividing by
the initial parton flux \( 2\hat{s} \), 
\begin{equation}
\label{x}
D^{jk}_{\mathrm{hard}}\! \left( \hat{s},b\right) =\frac{1}{2\hat{s}}2\mathrm{Im}\tilde{T}^{jk}_{\mathrm{hard}}(\hat{s},b),
\end{equation}
which gives

\begin{eqnarray}
D^{jk}_{\mathrm{hard}}\left( \hat{s},b\right)  & = & \frac{1}{8\pi ^{2}\hat{s}}\int d^{2}q_{\perp }\, \exp \! \left( -i\vec{q}_{\perp }\vec{b}\right) \, 2\mathrm{Im}\, T_{\mathrm{hard}}^{jk}(\hat{s},-q^{2}_{\perp })\\
 & = & \sigma _{\mathrm{hard}}^{jk}\! \left( \hat{s},Q_{0}^{2}\right) \frac{1}{4\pi R_{\mathrm{hard}}^{2}}\exp \! \left( -\frac{b^{2}}{4R_{\mathrm{hard}}^{2}}\right) \label{d-val-val} 
\end{eqnarray}

Since we also talk about ``valence-valence'' contribution, we use sometimes
\( D_{\mathrm{val}-\mathrm{val}} \) instead of \( D_{\mathrm{hard}} \):
\begin{equation}
\label{x}
D^{jk}_{\mathrm{val}-\mathrm{val}}\left( \hat{s},b\right) \equiv D^{jk}_{\mathrm{hard}}\left( \hat{s},b\right) ,
\end{equation}
 so these are two names for one and the same object.

\subsection{Semi-Hard Scattering}

Let us start discussing semi-hard sea-sea contribution, represented by a parton
ladder with ``soft ends'', see fig.\ \ref{fig:sea-sea}. As in case of soft
scattering, the external legs are nucleon constituents, connected to soft Pomerons.
The outer partons of the ladder are on both sides sea quarks or gluons (therefore
the index ``sea-sea''). The central part is exactly the hard scattering considered
in the preceding section. With a sum over all the hard scattering processes
in the center, we get the mathematical expression for the corresponding amplitude

\begin{equation}
iT_{\mathrm{sea}-\mathrm{sea}}(\hat{s},t)\! \! =\! \! \sum _{jk}\int ^{1}_{0}\! \frac{dz^{+}}{z^{+}}\frac{dz^{-}}{z^{-}}\, \, iT_{\mathrm{hard}}^{jk}(z^{+}z^{-}\hat{s},t)\, \, \mathrm{Im}\, T_{\mathrm{soft}}^{j}\! \! \left( \frac{s_{0}}{z^{+}},t\right) \, \mathrm{Im}\, T_{\mathrm{soft}}^{k}\! \! \left( \frac{s_{0}}{z^{-}},t\right) ,
\end{equation}
 with \( z^{\pm } \) being the momentum fraction of the external legs of the
parton ladder relative to the momenta of the nucleon constituents. The indices
\( j \) and \( k \) refer to the flavor of these external ladder partons.
The amplitudes \( T_{\mathrm{soft}}^{j} \) are the soft Pomeron amplitudes
discussed earlier, but with modified couplings, since the Pomerons are now connected
to the ladder on one side. The arguments \( s_{0}/z^{\pm } \) are the squared
masses of the two soft Pomerons, \( z^{+}z^{-}\hat{s} \) is the squared mass
of the hard piece.

Performing as usual the Fourier transform to the impact parameter representation
and dividing by \( 2\hat{s} \), we obtain the profile function
\begin{equation}
\label{x}
D_{\mathrm{sea}-\mathrm{sea}}\left( \hat{s},b\right) =\frac{1}{2\hat{s}}\, 2\mathrm{Im}\, \tilde{T}_{\mathrm{sea}-\mathrm{sea}}\! \left( \hat{s},b\right) ,
\end{equation}
 which may be written as
\begin{eqnarray}
D_{\mathrm{sea}-\mathrm{sea}}\left( \hat{s},b\right)  & = & \sum _{jk}\int ^{1}_{0}\! \! \! dz^{+}dz^{-}\! E_{\mathrm{soft}}^{j}\left( \! z^{+}\right) E_{\mathrm{soft}}^{k}\left( \! z^{-}\right) \sigma _{\mathrm{hard}}^{jk}(z^{+}z^{-}\hat{s},Q_{0}^{2})\label{d-sea-sea} \\
 & \times  & \frac{1}{4\pi \, \lambda ^{\! (2)}_{\mathrm{soft}}(1/(z^{+}z^{-}))}\exp \! \left( \! \! -\frac{b^{2}}{4\lambda ^{\! (2)}_{\mathrm{soft}}\! \left( 1/(z^{+}z^{-})\right) }\right) ,\nonumber \label{d-sea-sea} 
\end{eqnarray}
with the soft Pomeron slope \( \lambda ^{\! (2)}_{\mathrm{soft}} \) and the
cross section \( \sigma _{\mathrm{hard}}^{jk} \) being defined earlier. The
functions \( E_{\mathrm{soft}}^{j}\left( z^{\pm }\right)  \) representing the
``soft ends'' are defined as 
\begin{equation}
\label{x}
\mathrm{E}^{j}_{\mathrm{soft}}(z^{\pm })=\mathrm{Im}\, T_{\mathrm{soft}}^{j}\! \! \left( \frac{s_{0}}{z^{+}},t=0\right) .
\end{equation}
With the phenomenological description of soft Pomeron exchange\cite{gri68},
we parameterize the amplitude \( T_{\mathrm{soft}}^{\mathrm{g}} \) as 
\begin{equation}
\label{t-parton-gluon}
T_{\mathrm{soft}}^{\mathrm{g}}\, \, (\hat{s},t)=8\pi s_{0}\eta (t)\, \gamma _{\mathrm{part}}\gamma _{g}\, (\frac{\hat{s}}{s_{0}})^{\alpha _{\rm {p}}\, (0)}\exp \, (\lambda ^{\, (1)}_{\mathrm{soft}\, }(\hat{s}/s_{0})t)\, (1-\frac{s_{0}}{\hat{s}})^{\beta _{g}},
\end{equation}
 where 
\begin{equation}
\lambda ^{(1)}_{\mathrm{soft}}\! (z)=R_{\mathrm{part}}^{2}+\alpha '\! _{\mathrm{soft}}\ln \! z\, .
\end{equation}
So we get \( E_{\mathrm{soft}}^{\mathrm{g}} \):
\begin{equation}
\label{esoft-g-}
E_{\mathrm{soft}}^{g}\left( z\right) =8\pi s_{0}\gamma _{\mathrm{part}}\gamma _{g}\, z^{-\alpha _{\mathrm{soft}}\! (0)}\, (1-z)^{\beta _{g}}\, \, .
\end{equation}
With (\ref{esoft-g-}), we obtain \( E_{\mathrm{soft}}^{\mathrm{q}} \)
\begin{equation}
\label{esoft-q-}
E_{\mathrm{soft}}^{q}\left( z\right) =\gamma _{qg}\int ^{1}_{z}\! d\xi \, P_{g}^{q}(\xi )\, E_{\mathrm{soft}}^{g}\left( \frac{z}{\xi }\right) \, ,
\end{equation}
with 
\begin{equation}
\label{x}
\gamma _{qg}\gamma _{g}=w_{\mathrm{split}}\, \tilde{\gamma }_{g},\quad \gamma _{g}=\left( 1-w_{\mathrm{split}}\right) \, \tilde{\gamma }_{g},
\end{equation}
 where \textbf{\textit{\( w_{\mathrm{split}} \)}} parameter determined the
relative weight of sea quark content of the soft Pomeron (the probability for
\( g\rightarrow q\bar{q} \) splitting in the soft Pomeron). By the momentum
conservation constraint 
\begin{equation}
1=\int _{0}^{1}\, dz\, \sum _{j}\, z\, E_{\mathrm{soft}}^{\mathrm{j}}\, (z)=8\pi s_{0}\gamma _{\mathrm{part}}\widetilde{\gamma }_{g}\int _{0}^{1}\, dz\, z^{1-\alpha _{\mathrm{soft}}(0)}(1-z)^{\beta _{g}}\, \, ,
\end{equation}
 \( \tilde{\gamma }_{g} \) is fixed as 
\begin{equation}
\label{x}
\tilde{\gamma }_{g}=\frac{1}{8\pi s_{0}\gamma _{\mathrm{part}}}\frac{\Gamma \! \left( 3-\alpha _{\mathrm{soft}}\! (0)+\beta _{g}\right) }{\Gamma \! \left( 2-\alpha _{\mathrm{soft}}\! (0)\right) \, \Gamma \! \left( 1+\beta _{g}\right) }.
\end{equation}
We neglected the small hard scattering slope \( R_{\mathrm{hard}}^{2} \) compared
to the Pomeron slope \( \lambda _{\mathrm{soft}} \). We call \( E_{\mathrm{soft}} \)
also the `` soft evolution'', to indicate that we consider this as simply
 a continuation of the QCD evolution, however, in a region where perturbative
 techniques do not apply any more. It's easy to see that \( E_{\mathrm{soft}}^{j}\left( z\right)  \)
has the meaning of the momentum distribution of parton \( j \) in the soft
Pomeron at virtuality \( Q_{0}^{2} \). 

Consistency requires to also consider the mixed semi-hard contributions with
a valence quark on one side and a sea quark or gluon on the other one, see fig.\ \ref{fig:mixed}.
We have
\begin{equation}
\label{x}
iT^{j}_{\mathrm{val}-\mathrm{sea}}(\hat{s},t)=\int ^{1}_{0}\! \frac{dz^{-}}{z^{-}}\sum _{k}\mathrm{Im}\, T_{\mathrm{soft}}^{k}\! \! \left( \frac{s_{0}}{z^{-}},t\right) iT_{\mathrm{hard}}^{jk}\left( z^{-}\hat{s},t\right) \qquad 
\end{equation}
and
\begin{equation}
\label{d-val-sea}
D^{j}_{\mathrm{val}-\mathrm{sea}}\left( \hat{s},b\right) =\sum _{k}\int ^{1}_{0}\! dz^{-}\, E_{\mathrm{soft}}^{k}\left( z^{-}\right) \, \sigma _{\mathrm{hard}}^{jk}\! \left( z^{-}\hat{s},Q_{0}^{2}\right) \, \frac{1}{4\pi \, \lambda ^{\! (1)}_{\mathrm{soft}}(1/z^{-})}\exp \! \left( \! \! -\frac{b^{2}}{4\lambda ^{\! (1)}_{\mathrm{soft}}\! \left( 1/z^{-}\right) }\right) 
\end{equation}
 where \( j \) is the flavor of the valence quark at the upper end of the ladder
and \( k \) is the type of the parton on the lower ladder end. Again, we neglected
the hard scattering slope \( R^{2}_{\mathrm{hard}} \) compared to the soft
Pomeron slope. A contribution \( D^{j}_{\mathrm{sea}-\mathrm{val}}\left( \hat{s},b\right)  \),
corresponding to a valence quark participant from the target hadron, is given
by the same expression,
\begin{equation}
\label{x}
D^{j}_{\mathrm{sea}-\mathrm{val}}\left( \hat{s},b\right) =D^{j}_{\mathrm{val}-\mathrm{sea}}\left( \hat{s},b\right) ,
\end{equation}
since eq.\ (\ref{d-val-sea}) stays unchanged under replacement \( z^{-}\rightarrow z^{+} \)
and only depends on the total c.m. energy squared \( \hat{s} \) for the parton-parton
system.

\subsection{Nucleon-Nucleon Scattering}

Let us define for any of the elementary interaction types \( K \) (soft, hard,
sea-sea, sea-val, and val-sea) a ``dressed'' partonic profile function \( G \)
via
\begin{eqnarray}
G_{K}(x^{+},x^{-},s,b)\! \! \!  & = & \! \! \! \frac{1}{2x^{+}x^{-}s}2\mathrm{Im}\, \tilde{T}'\! _{K}(x^{+},x^{-},s,b),\label{gss} 
\end{eqnarray}
 where \( \tilde{T}'\! _{K} \) is the Fourier transform of \( T'\! _{K} \),
with
\begin{equation}
\label{thh}
T'\! _{K}\left( x^{+},x^{-},s,q^{2}\right) \! \! =\! \! T_{K}\left( x^{+}x^{-}s,q^{2}\right) F_{K^{+}}(x^{+})F_{K^{-}}(x^{-})\; \exp \! \left( 2R_{N}^{2}\, q^{2}\right) 
\end{equation}
representing the contributions of ``elementary interactions plus external legs''.
Here we assumed a simple factorized form for the vertex for nucleon coupling
to \( n \) constituent ``soft'' partons participating in elementary scattering
processes (external partons for processes of fig.\ \ref{fig:val-val}, \ref{fig:sea-sea},
\ref{fig:mixed}, \ref{fig:soft}) :
\begin{equation}
\label{N-vertex}
N_{N}^{(n)}\! \left( p,k_{1},\ldots ,k_{n},q_{1},\ldots ,q_{n}\right) =\prod _{i=1}^{n}\left[ F_{K_{i}^{\pm }}\left( x_{i}^{\pm }\right) \, \exp \! \left( R_{N}^{2}\, \sum ^{n}_{k=1}q_{i}^{2}\right) \right] \, F_{\mathrm{remn}}\left( 1-\sum ^{n}_{j=1}x^{\pm }_{j}\right) ,
\end{equation}
 where \( x^{\pm }=k^{\pm }/p^{\pm } \) are the light cone momentum fractions
of the \( i \)-th constituent parton (\( \pm  \) correspond to the projectile/target
case), \( q_{i} \) is the 4-momentum transfer in the \( i \)-th scattering
process, \( K^{\pm }_{i} \) is the type of \( i \)-th constituent parton (\( K^{\pm }_{i}= \)``sea''
for the ``soft'' or ``sea-sea'' type process; \( K^{\pm }_{i}= \)``val''
for the ``hard'' scattering; \( K^{+}_{i}= \)``val'', \( K^{-}_{i}= \)``sea''
for the ``val-sea'' case etc.), and \( R_{N}^{2} \) is known as the nucleon
Regge slope. The factors in the square brackets are then associated with individual
scattering contributions and included in the definition of \( T'\! _{K},G_{K} \).

For the functions \( F_{\mathrm{sea}},\, F_{\mathrm{remn}} \) we use a simple
Regge-inspired ansatz:
\begin{equation}
\label{f-part-main}
F_{\mathrm{sea}}(x)=\gamma _{N}x^{-\alpha _{\mathrm{part}}},
\end{equation}
\begin{equation}
\label{f-remn-main}
F_{\mathrm{remn}}\! (x)=x^{\alpha _{\mathrm{remn}}}.
\end{equation}
where the different parameters \( \alpha  \) are finally determined by comparing
with experimental data. The function \( F_{\mathrm{val}} \) is constructed
such that it reproduces an input parametrization (GRV94 \cite{glu95}) for valence
quark momentum distributions in the nucleon \( q_{v}\! \left( x,Q_{0}^{2}\right)  \)
at the initial scale \( Q_{0}^{2} \) .

Based on the above definitions, we may write 
\begin{equation}
\label{ghh}
G_{K}\left( x^{+},x^{-},s,b\right) =D'\! _{K}\left( x^{+}x^{-}s,b\right) \, F_{K^{+}}(x^{+})\, F_{K^{-}}(x^{-}),
\end{equation}
with the \( D' \)-functions expressed in terms of the bare parton profile functions
as 
\begin{equation}
D'\! _{K}(\hat{s},b)=\int \! d^{2}b'\, D_{K}(\hat{s},|\vec{b}-\vec{b}'|)\, \frac{1}{8\pi R^{2}_{N}}\, \exp \! \left[ -\frac{b'^{2}}{8R^{2}_{N}}\right] ,
\end{equation}
which means that \( D'\! _{K} \) has the same functional form as \( D_{K} \),
with \( \lambda ^{(n)}\! (\xi ) \) being replaced by \( \lambda ^{(n)}(\xi )+2R^{2}_{N} \).

So we have the ``bare'' partonic profile functions \( D_{K} \) and the ``dressed''
ones \( G_{K} \), where the former ones describe pure parton-parton \cite{hla01}
scattering (or more generally scattering between partonic constituents), whereas
the latter ones describe partonic scattering, however, taking into account the
longitudinal and transverse momentum distribution of the partons in the nucleons.
These dressed profile functions are the elementary quantities, based on which
one may construct a proper multiple scattering theory, for nucleon-nucleon as
well as nucleus-nucleus scattering.

Before we discuss a more correct treatment, let us first show how we recover
the conventional Gribov-Regge approach of multiple scattering. Here one considers
so-called eikonals,
\begin{equation}
\chi _{K}(s,b)=\int dx^{+}dx^{-}G_{K}(x^{+},x^{-},s,b)\, F_{\mathrm{remn}}\! \! \left( 1-\! x^{+}\right) \, F_{\mathrm{remn}}\! \! \left( 1-\! x^{-}\right) ,
\end{equation}
and using the eikonal approximation, one may obtain the inelastic cross section
as
\begin{equation}
\sigma _{\mathrm{inel}}(s)=\int d^{2}b\left\{ 1-\exp \left[ -\sum _{K}\chi _{K}(s,b)\right] \right\} .
\end{equation}
 To obtain such a simple formula, one has to ignore the fact that the energy
has to be shared between the different elementary interactions, which is certainly
not correct.

Doing energy conservation properly, one obtains \cite{hla01}
\begin{eqnarray}
\sigma _{\mathrm{inel}}(s) & = & \int d^{2}b\sum ^{\infty }_{m=1}\frac{1}{m!}\, \int ^{1}_{0}\! \prod ^{m}_{\mu =1}\! dx_{\mu }^{+}dx_{\mu }^{-}\prod ^{m}_{\mu =1}G(x_{\mu }^{+},x_{\mu }^{-},s,b)\, \sum ^{\infty }_{l=0}\frac{1}{l!}\, \int ^{1}_{0}\! \prod ^{l}_{\lambda =1}\! d\tilde{x}_{\lambda }^{+}d\tilde{x}_{\lambda }^{-}\nonumber \\
 & \times  & \prod ^{l}_{\lambda =1}-G(\tilde{x}_{\lambda }^{+},\tilde{x}_{\lambda }^{-},s,b)\, F_{\mathrm{remn}}\left( x^{\mathrm{proj}}\! \! -\! \! \sum _{\lambda }\tilde{x}_{\lambda }^{+}\right) F_{\mathrm{remn}}\left( x^{\mathrm{targ}}\! \! -\! \! \sum _{\lambda }\tilde{x}_{\lambda }^{-}\right) ,\label{gam-agk-g} 
\end{eqnarray}
with \textbf{\textit{\( x_{\mu }^{\pm },x_{\lambda }^{\pm } \)}} being the
light cone momentum fractions of constituent partons for the elementary scattering
contributions and
\begin{equation}
x^{\mathrm{proj}/\mathrm{targ}}=1-\sum x^{\pm }_{\mu }
\end{equation}
being the momentum fraction of the projectile/target remnant, and with 
\begin{equation}
\label{g1p-tot}
G=G_{\mathrm{soft}}+G_{\mathrm{val}-\mathrm{val}}+G_{\mathrm{sea}-\mathrm{sea}}+G_{\mathrm{val}-\mathrm{sea}}+G_{\mathrm{sea}-\mathrm{val}}
\end{equation}
being the sum of all the contributions, soft ones, hard ones, and semi-hard
ones. The terms \( +G \) in formula (\ref{gam-agk-g}) correspond to cut elementary
diagrams, the terms \( -G \) to the uncut ones, the sums represent summations
over all possible cuts. Eq. (\ref{gam-agk-g}) can be written as
\begin{equation}
\label{sig}
\sigma _{\mathrm{inel}}(s)=\int d^{2}b\left\{ 1-\Phi (s,b)\right\} ,
\end{equation}
with 
\begin{equation}
\label{phi}
\Phi (s,b)=\sum ^{\infty }_{l=0}\frac{1}{l!}\, \int ^{1}_{0}\! \prod ^{l}_{\lambda =1}\! d\tilde{x}_{\lambda }^{+}d\tilde{x}_{\lambda }^{-}\prod ^{l}_{\lambda =1}-G(\tilde{x}_{\lambda }^{+},\tilde{x}_{\lambda }^{-},s,b)\, F_{\mathrm{remn}}\left( 1-\! \! \sum _{\lambda }\tilde{x}_{\lambda }^{+}\right) F_{\mathrm{remn}}\left( 1-\! \! \sum _{\lambda }\tilde{x}_{\lambda }^{-}\right) ,
\end{equation}
which can be evaluated numerically. This energy sharing formalism can be easily
generalized to nucleus-nucleus scattering \cite{hla01}, not only for calculating
cross sections, but also particle production.

\section{Some Results and Discussion}

\subsection{The Behavoir of the ``Dressed Profile Function'' \protect\( G\protect \) }

\begin{figure}[htb]
{\par\centering \resizebox*{!}{0.24\textheight}{\includegraphics{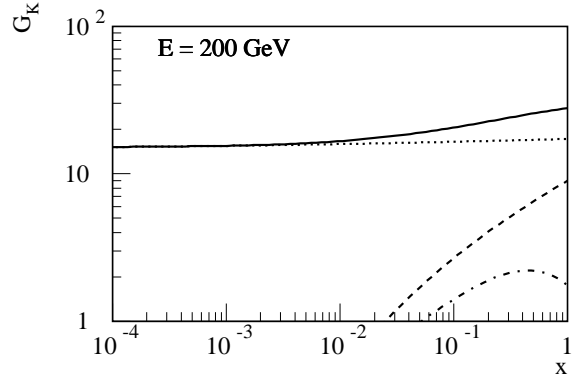}} \par}

\caption{The functions \protect\( G_{K}\protect \) in dependence on \protect\( x=x^{+}x^{-}\protect \)at
\protect\( b=0\protect \) for the different contributions \protect\( K\protect \)
for a cms energy of 200 GeV. {\small }We show soft (dotted), semi-hard (dashed),
and valence contributions (dashed-dotted), where semi-hard is meant to be the
sum of sea-sea, sea-val, and val-sea. \label{fig:g200}}
\end{figure}

\begin{figure}[htb]
{\par\centering \resizebox*{!}{0.24\textheight}{\includegraphics{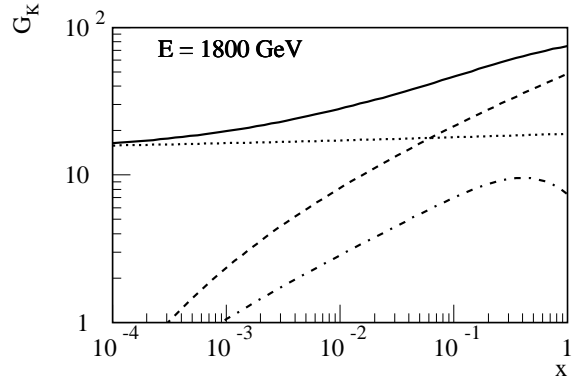}} \par}

\caption{The functions \protect\( G_{K}\protect \) in dependence on \protect\( x=x^{+}x^{-}\protect \)at
\protect\( b=0\protect \) for the different contributions \protect\( K\protect \)
for a cms energy of 1800 GeV. {\small }We show soft (dotted), semi-hard (dashed),
and valence contributions (dashed-dotted), where semi-hard is meant to be the
sum of sea-sea, sea-val, and val-sea.\label{fig:g1800}}
\end{figure}

As discussed in detail in the preceding chapter, one may express the inelastic
cross section (and many other quantities) in terms of the ``dressed profile
function'' \( G \), which is a sum of soft, hard, and semi-hard contributions:
\( G=\sum G_{K} \). This function can be interpreted as the number of elementary
interactions (``Pomerons'') with light-cone momentum fractions \( x^{+} \)
and \( x^{-} \) at a given impact parameter \( b \). We first investigate
the functional form of these functions \( G_{K} \). We plot in figs. \ref{fig:g200}
and \ref{fig:g1800} the functions \( G_{K} \) 
in dependence on \( x=x^{+}x^{-} \)at \( b=0 \) for a cms energy of 200 and
1800 GeV for the different contributions \( K \) (we consider the sum of sea-sea,
sea-val, and val-sea, referred to as ``semi-hard''). The dominant contributions
are in any case the soft and the semi-hard ones. Both show roughly a power-law
behavior, with a much steeper increase of the semi-hard component. At 200 GeV,
the soft component is by far dominant, whereas for 1800 GeV the semi-hard component
is taking over, however, with the soft one still being bigger at small \( x \).
Although the semi-hard part has a cutoff at some small value of \( x \), the
complete contribution shows always a smooth and regular behavior in the limit
of \( x \) going to zero, due to the soft component. 

The smooth behavior of the \( G \) function has also the very nice side effect
that we can approximate it (to a very good precision) by a simple analytical
function, which allows an analytical calculation of many interesting quantities,
in particular in nucleus-nucleus collisions, where numerical calculations are
quite costly, or even impossible.

\begin{figure}[htb]
{\par\centering \resizebox*{!}{0.26\textheight}{\includegraphics{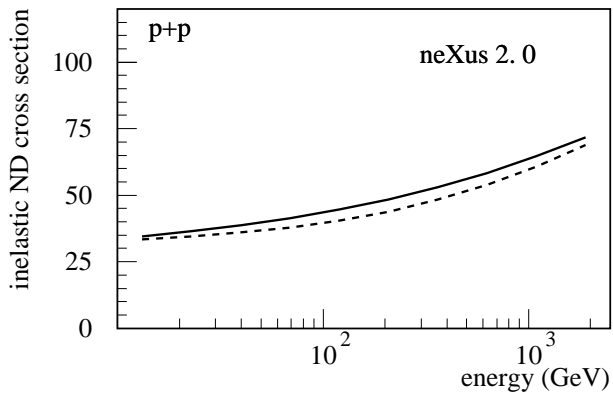}} \par}

\caption{The inelastic cross section for proton-proton scattering as a function of the
energy \protect\( \sqrt{s}\protect \). \label{fig:sigma}}
\end{figure}
 Having calculated \( G \), we can now proceed to calculate the inelastic cross
section, using eqs. (\ref{sig},\ref{phi}). We obtain the result shown in fig.
\ref{fig:sigma} as a full curve. Accounting for lowest order screening corrections
due to so-called enhanced Pomeron diagrams we obtained somewhat reduced values
of \( \sigma _{\mathrm{inel}}(s) \), shown in fig. \ref{fig:sigma} as the
dashed curve.

\subsection{Comparison with Data: }

\begin{figure}[htb]
{\par\centering \resizebox*{!}{0.26\textheight}{\includegraphics{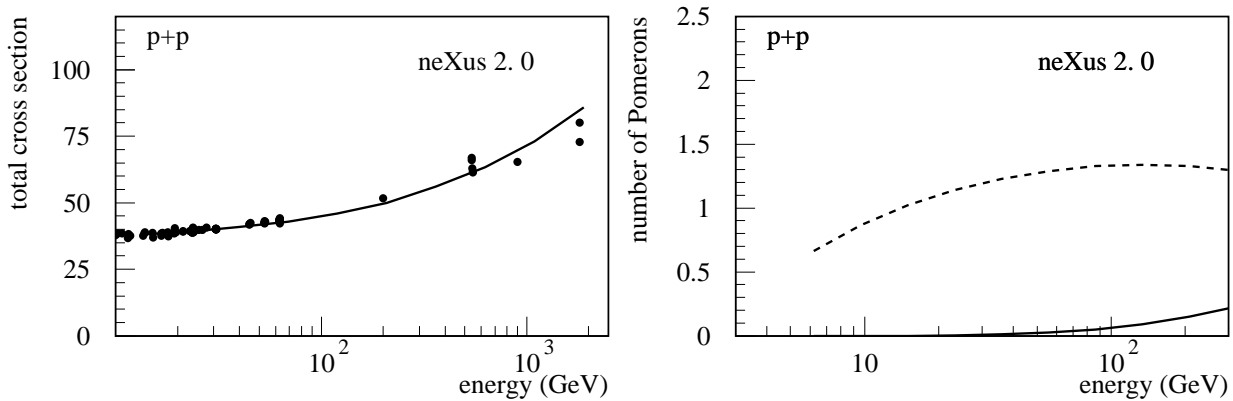}} \par}

\caption{The total cross section as a function of the energy \protect\( \sqrt{s}\protect \)
(left figure): the full line is the simulation, the points represent data \cite{cas98}.
Pomeron numbers as a function of energy \protect\( \sqrt{s}\protect \) (right):
soft(dashed) and semi-hard (solid line) Pomerons. \label{fig-his1}}
\end{figure}

\begin{figure}[htb]
{\par\centering \resizebox*{!}{0.5\textheight}{\includegraphics{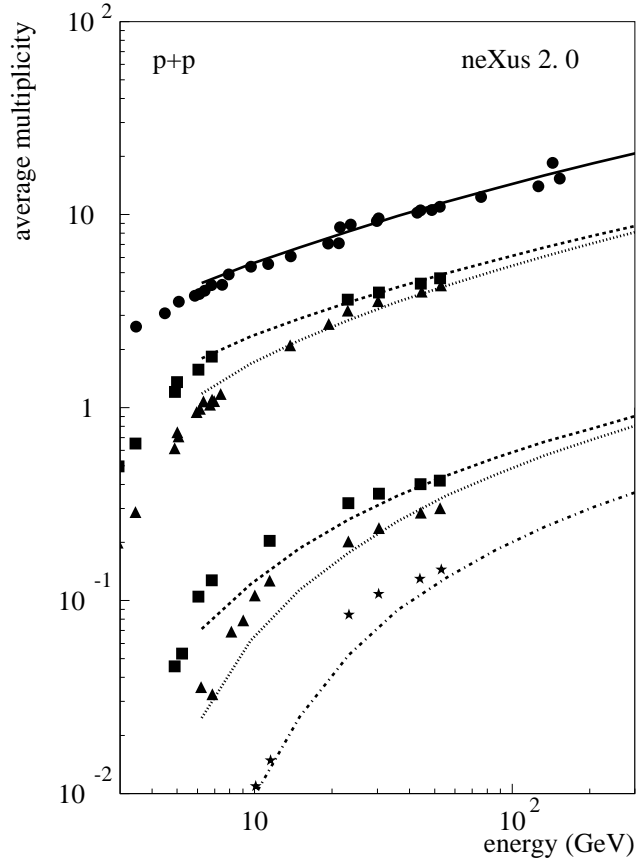}} \par}

\caption{\label{fig-his2}The average multiplicities of different hadron species, as
a function of the energy \protect\( \sqrt{s}\protect \). From top to bottom:
all charged particles, \protect\( \pi ^{+}\protect \), \protect\( \pi ^{-}\protect \),
\protect\( \rm {K}^{+}\protect \),\protect\( \rm {K}^{-}\protect \),\protect\( \overline{p}\protect \).
The full lines are simulations, the points represent dada from \cite{gia79}.}
\end{figure}

An important advantage of our approach is that it allows not only to calculate
different interaction cross sections but also to develop a fully self-consistent
Monte Carlo procedure to simulate hadron-hadron (nucleus-nucleus) interactions,
including direct modeling of perturbative parton evolution in hard and semihard
elementary processes. Thus, we were able to simulate many processes and obtain
valuable results which agree quite well with experimental data. Here we present
our results for proton-proton interactions in the energy range between roughly
10 and 2000 GeV, which represents the range of validity of our approach. The
energy dependence of the total cross section has been compared with the data
\cite{cas98} in figure \ref{fig-his1}. The average multiplicities of different
hadron species as a function of the energy \( \sqrt{s} \) are shown in figure
\ref{fig-his2}.

\begin{figure}[htb]
{\par\centering \resizebox*{!}{0.26\textheight}{\includegraphics{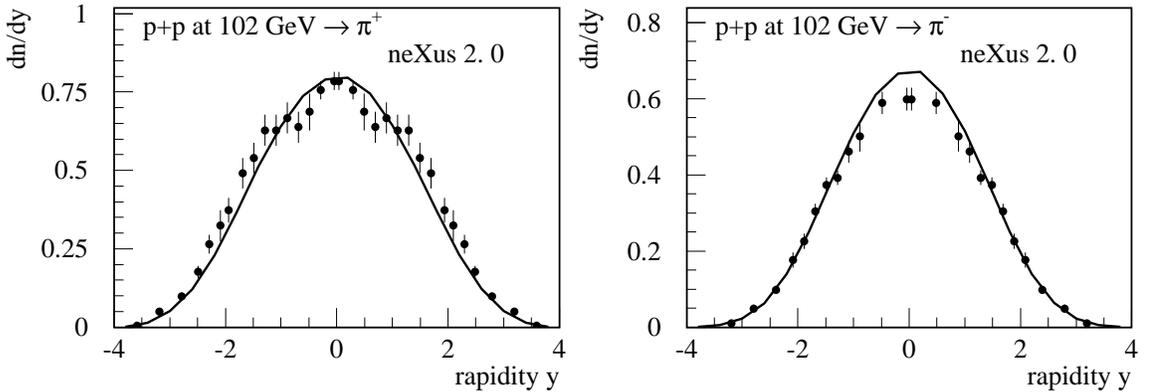}} \par}

\caption{\label{fig-his4}Rapidity distributions of pions at 100 GeV. The full lines
are simulations, the points represent data \cite{whi74}.}
\end{figure}
\begin{figure}[htb]
{\par\centering \resizebox*{!}{0.42\textheight}{\includegraphics{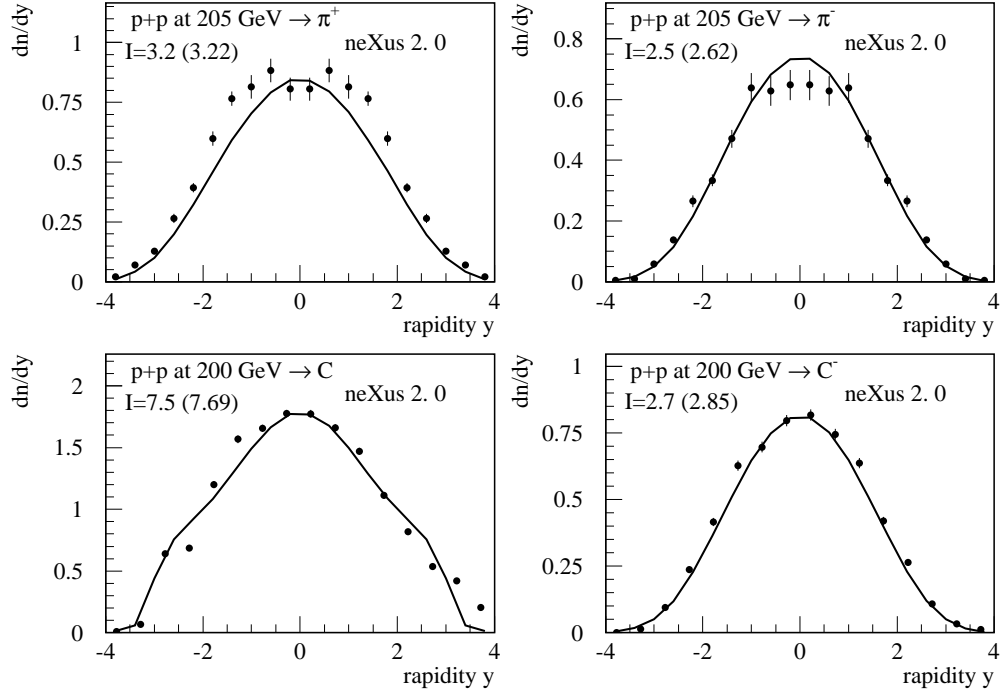}} \par}

\caption{\label{fig-his5}Psedo-rapidity distributions of pions (\protect\( \pi ^{+}\protect \),\protect\( \pi ^{-}\protect \))and
charged particles (all charged and negatively charged) at 200 GeV. The full
lines are simulations, the points represent data \cite{whi74,dem82}. }
\end{figure}
\begin{figure}[htb]
{\par\centering \resizebox*{!}{0.24\textheight}{\includegraphics{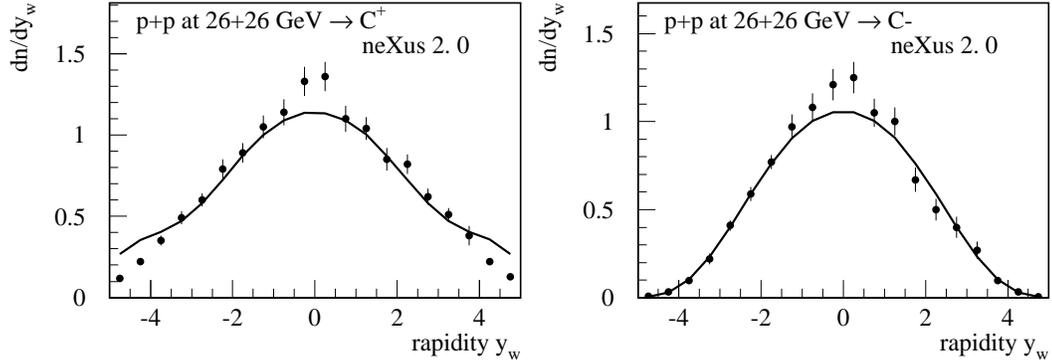}} \par}

\caption{\label{fig-his6}Psedo-rapidity distributions of positively and negatively
charged particles at 53 GeV (cms). The full lines are simulations, the points
represent data \cite{bre83}.}
\end{figure}
\begin{figure}[htb]
{\par\centering \resizebox*{!}{0.245\textheight}{\includegraphics{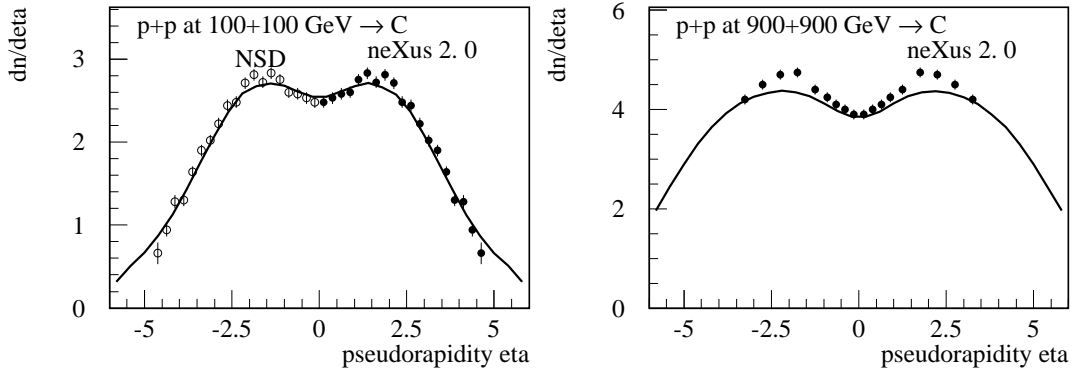}} \par}

\caption{\label{fig-his7}Psedo-rapidity distributions of charged particles at 200 and
1800 GeV (cms). The full lines are simulations, the points represent data \cite{aln86,abe90}.}
\end{figure}
 
\begin{figure}[htb]
{\par\centering \resizebox*{!}{0.245\textheight}{\includegraphics{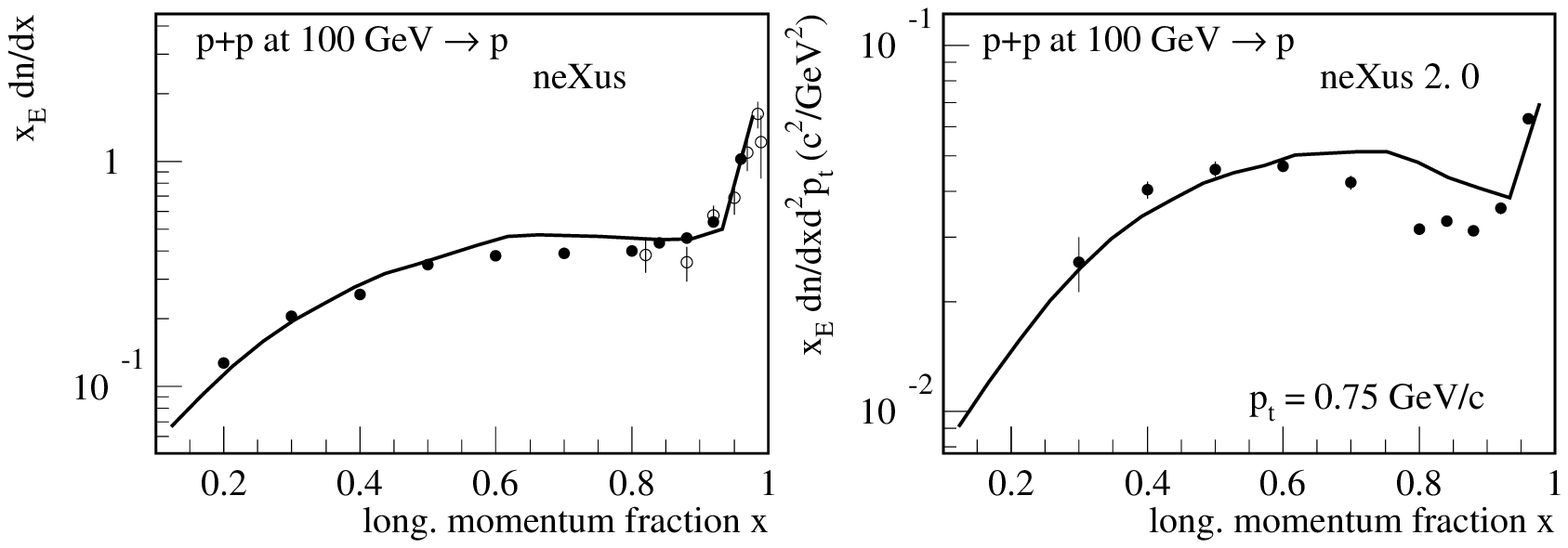}} \par}

\caption{\label{fig-his11}Longitudinal momentum fraction distribution of protons at
100 GeV, integrated over \protect\( p_{t}\protect \) (left) and for \protect\( p_{t}=0.75\protect \)
GeV/c (right). The full lines are simulations, the points represent data \cite{bre82}.}
\end{figure}
 
\begin{figure}[htb]
{\par\centering \resizebox*{!}{0.245\textheight}{\includegraphics{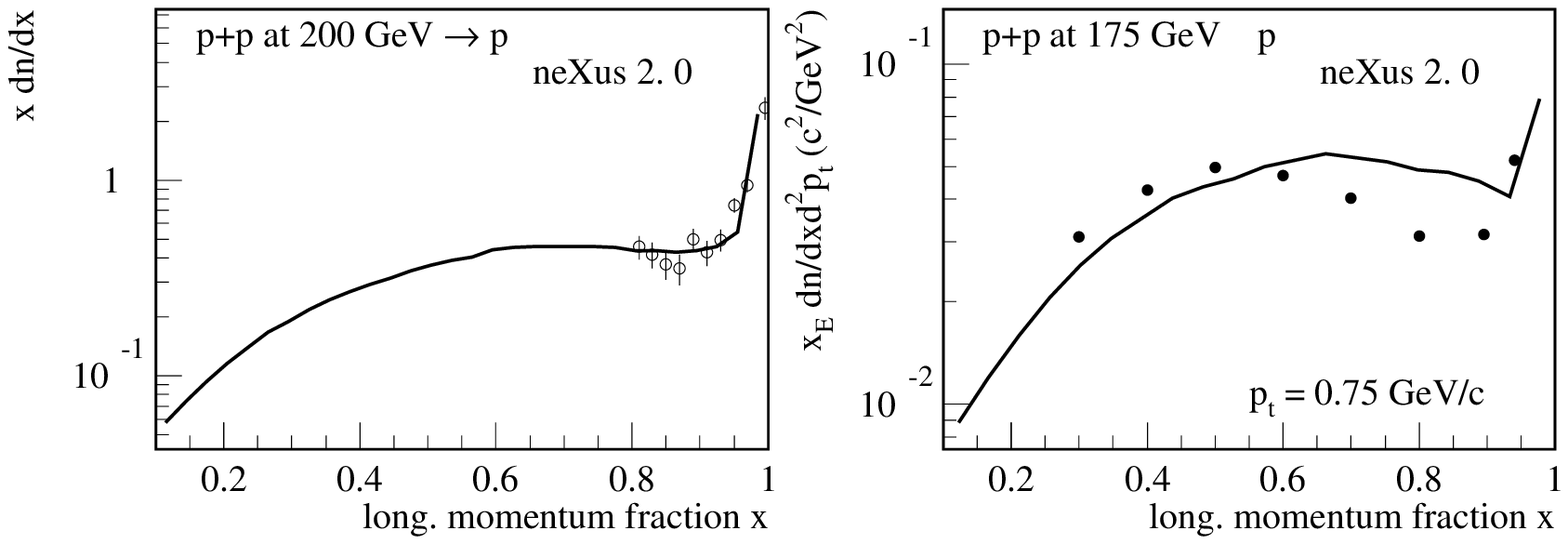}} \par}

\caption{\label{fig-his12}Longitudinal momentum fraction distribution of protons at
200 GeV, integrated over \protect\( p_{t}\protect \) (left) and at 175 GeV
for \protect\( p_{t}=0.75\protect \) GeV/c (right). The full lines are simulations,
the points represent data \cite{whi74,bre82}.}
\end{figure}
 We also present the spectrums of pion here. Figure \ref{fig-his4} shows the
rapidity distributions of pions at 100 GeV and figure \ref{fig-his5} shows
the psedo-rapidity distributions of pions (\( \pi ^{+} \),\( \pi ^{-} \))and
charged particles (all charged and negatively charged) at 200 GeV. Figure \ref{fig-his6}
shows the psedo-rapidity distributions of positively and negatively charged
particles at 53 GeV (cms). Figure \ref{fig-his7} shows the psedo-rapidity distributions
of charged particles at 200 and 1800 GeV (cms). 

The spectrums of protons are presented, too. Figure \ref{fig-his11} shows the
longitudinal momentum fraction distribution of protons at 100 GeV, integrated
over \( p_{t} \) (left) and for \( p_{t}=0.75 \) GeV/c (right). Figure \ref{fig-his12}
shows the longitudinal momentum fraction distribution of protons at 200 GeV,
integrated over \( p_{t} \) (left) and at 175 GeV for \( p_{t}=0.75 \) GeV/c
(right).

\newpage
\subsection{Some Discussion}

This paper has shown a theoretical discussion of a new concept of treating soft
and hard scattering in a consistent fashion. As discussed above, at energies
presently (or in the near future) accessible, both soft and hard components
play an important role, so one cannot simply ignore any of the two. This was
our main motivation to develop the theoretical framework discussed in this paper.
The full self-consistent scheme for the description of high energy hadron-hadron
(nucleus-nucleus) interactions has to account also for all significant screening
corrections, or more generally interactions between elementary diagrams. This
part of the work will be discussed elsewhere \cite{ost01}.

\bibliographystyle{pr2}
\bibliography{a}

\end{document}